# Analysis of Acoustic Nonlinearity Parameter B/A in Liquids Containing Ultrasound Contrast Agents

Lang Xia

Email: lang4xia@gmail.com

**Abstract**

The acoustic nonlinearity parameter B/A plays a significant role in the characterization of acoustic properties of various biomaterials and biological tissues. It has the potential to be a favorable imaging modality in contrast ultrasound imaging with coated microbubbles. However, the development of effective means for evaluating the nonlinearity parameter of suspensions of ultrasound contrast agents (UCAs, also known as bubbly liquids) remains open. The present paper formulates a new equation based on the thermodynamic method that correlates both attenuation and phase velocity of linear ultrasound. The simplicity of the present method makes the B/A estimation possible with a relatively rigorous mathematical derivation. The calculated nonlinearity parameter contains the contribution of dynamic effects of bubbles, and its low-frequency limit agrees with B/A estimated by the method of mixture law when the volume fraction is below $10^{-4}$. Furthermore, the maximum B/A in bubbly liquids can reach up to $10^5$, while the minimum can be as low as $-10^5$. The negative nonlinearity parameter indicates significantly different thermodynamic properties of bubbly liquids.

## I. INTRODUCTION

Due to the inherent connection between the nonlinearity parameter B/A and the acoustic property of materials, B/A has been an important physical constant of characterizing different acoustic materials and biological media (Law, Frizzell et al. 1985, Errabolu, Sehgal et al. 1988, Dong, Madsen et al. 1999, Duck 2002, Gong, Zhang et al. 2004). B/A imaging has also gained much attention recently (Varray, Basset et al. 2011, Van Sloun, Demi et al. 2015). The nonlinearity parameter B/A is a ratio of expansion coefficients derived from the equation of state of fluids (pressure and density relation) for an adiabatic process by the Taylor series expansion up to the second order (Fox and Wallace 1954, Beyer 1960). Since Beyer proposed B/A as a material constant for acoustic media, the nonlinearity parameter of various materials, including industrial, chemical and biological fluids, has been studied and measured by a thermodynamic technique (Bjørnø 1982, Bjørnø 2002, Trarieux, Callé et al. 2014, Renaud, Bosch et al. 2015).

Microbubbles improve the signal to noise ratio in acoustic measurements due to the compressibility of the gas core. Coated microbubbles have been used as ultrasound contrast agents for decades in contrast ultrasound imaging. The presence of bubbles in a liquid introduces dispersion, increased attenuation, and nonlinearities to the medium (bubbly liquid). The dispersion relation and attenuation of acoustic waves—that provide a feasible way to characterize the rheological properties of the bubbles (or Ultrasound Contrast Agents) —have been studied extensively both in theories and experiments (Wijngaarden 1972, Commander and Prosperetti 1989, Sarkar, Shi et al. 2005, Xia, Porter et al. 2015). The development of effective means for evaluating B/A of bubbly liquids has not yet received sufficient attention, regardless of the importance of the second-order acoustic nonlinearity parameter. Currently, three major methods are commonly seen in the estimation of the nonlinearity parameter of bubbly liquids. The thermodynamic method estimates B/A from the relation of phase velocity and hydrostatic pressure (Hélène, Anthony et al. 2009, Renaud, Bosch et al. 2015). The finite-amplitude method correlates the B/A to the ratio of the second harmonic and source amplitudes of the pressure wave due to the medium-induced wave distortion and then the transfer of energy at the fundamental frequency to other harmonics (Beyer 1960, Cobb 1983). Beyer was the first who proposed the finite-amplitude method for measuring B/A of different materials. Cobb used an improved finite-amplitude method to deal with dissipation effects of acoustic waves in the medium (Cobb 1983). The expression in his paper was then adopted by many authors to measure B/A of bubbly liquids (Wu, Zhu et al. 1995, Ma, Yu et al. 2004). Another important method is the mixture law of nonlinearity parameters (Apfel 1983, Everbach, Zhu et al. 1991). This method correlates the effective nonlinearity parameter of immiscible fluids to their mass fractions. It may only be suitable to measure bubbly liquids at low-frequency limit (Zaitsev, Dyskin et al. 2009). Bjørnø also proposed the so-called parametric-array method to characterize the nonlinearity parameters B/A of bubbly liquids (Bjørnø 1982). This method is based on the equation derived by Westervelt using parametric array (Westervelt 1963). The effective B/A was obtained as a result of the coefficient comparison. It is able to capture B/A in dispersive media at low bubble volume fractions.

In experimental setting, the thermodynamic method calculates B/A directly by varying hydrostatic pressures to determine the changes of sound speed (Beyer 1960, Trarieux, Callé et al. 2014, Renaud, Bosch et al. 2015), the finite-amplitude method relates B/A to the amplitude of the second harmonics (Wu, Zhu et al. 1995), and the method of mixture law correlates the B/A to the volume fraction of bubbles. While analytical formulae for calculating B/A using the method of mixture law are well accepted, effective equations for determining the nonlinearity parameter in bubbly liquids using the finite-amplitude method remains elusive, and analytical methods originated directly from the thermodynamic approach is still missing. We thus develop a method to investigate B/A of bubbly liquids based on the thermodynamic approach. The current method relies on the dispersion relation of the bubbly liquids. The derived formulae in the present paper enable us to estimate B/A from the attenuation coefficient and phase velocity of acoustic waves in the media. The nonlinearity parameter B/A is calculated for suspensions containing polydisperse, monodisperse, and coated microbubbles. The scope of the current method is also discussed.

## II. THEORETICAL BACKGROUND

For the convenience of reading, we briefly review the existing methods for the evaluation of B/A in bubbly liquids and address possible concerns for each of them. Noticing that a lot of detailed treatments of nonlinear acoustics are available in the literature, we limit ourselves to work relevant only to bubbly liquids. Hence, this review may be subjected to some sorts of bias.

### II A. Thermodynamic method

The earliest literature covering nonlinearity in an acoustic medium distorting the propagating waves may date back to Earnshaw and Riemann (Earnshaw 1860), who correlated the distortion of sound waves to the polytropic constant of a gas. In order to characterize the nonlinearity in fluids, Fox and Wallace first gave the definition of B/A that is equal to $\kappa-1$ when the medium is purely an ideal gas (Fox and Wallace 1954). From this perspective, B/A of a liquid is analogous to the polytropic constant of a gas, acoustically. The expression of B/A was obtained by assuming the pressure is a function of density and entropy $P = P(\rho, s)$ and applying the Taylor expansion at $\rho = \rho_0$ to the relation for the isentropic process (Fox and Wallace 1954, Beyer 1960)

$$P = P_0 + A\frac{\rho - \rho_0}{\rho_0} + \frac{B}{2}\left(\frac{\rho - \rho_0}{\rho_0}\right)^2 + \cdots$$

where

$$A := \rho_0 \left(\frac{\partial P}{\partial \rho}\right)_{s, \rho = \rho_0} = \rho_0 c_0^2$$

$$B := \rho_0^2 \left(\frac{\partial^2 P}{\partial \rho^2}\right)_{s, \rho = \rho_0}$$

The subscript $s$ indicates that the entropy is to be held constant in the process. $c_0$ is the sound speed in the host liquid. The nonlinearity parameter is defined as

$$\frac{B}{A} = \frac{\rho_0}{c_0^2}\left(\frac{\partial^2 P}{\partial \rho^2}\right)_{s, \rho = \rho_0} = \frac{\rho_0}{c_0^2}\frac{\partial}{\partial \rho}\left(\frac{\partial P}{\partial \rho}\right)_{s, \rho = \rho_0} = 2\rho_0 c_0 \left(\frac{\partial V_p}{\partial P}\right)_{s, P = P_0}$$

where the $V_p$ is the instantaneous sound speed (or phase velocity for dispersive media). Since the nonlinearity of a medium is closely related to the molecular structure and heterogeneity of the medium, Beyer proposed the nonlinearity parameter B/A could be employed in the characterization of biological fluids and soft tissues (Beyer 1960). Experimental measurements of B/A may be performed by varying the hydrostatic pressure and then measuring the corresponding changes of the phase velocity with the following approximation

$$\frac{B}{A} \approx 2\rho_0 c_0 \left(\frac{\Delta V_p}{\Delta P}\right)_{s, P = P_0}$$

where $\Delta V_p = V_{p1} - V_{p2}$ and $\Delta P = P_1 - P_2$.

This method requires the pressure changing rapidly to maintain the isentropic condition but still smoothly to avoid the formation of shocks. As for the measurement of dispersive media, the use of a monochromatic wave usually ignores a fact that the difference of phase velocity may not be appreciable with respective to the hydrostatic pressure. Therefore, experimental measurements based on the above method require elaborate techniques and apparatus (Law, Frizzell et al. 1985, Cain 1986, Trivett, Pincon et al. 2006, Hélène, Anthony et al. 2009, Trarieux, Callé et al. 2014), which may not be suitable in the standard ultrasound exams (Varray, Basset et al. 2011). This concern is even prominent when it comes to bubbly liquids as they are highly dispersive, and the variation of phase velocity is closely associated with the frequency of propagating waves. On the other hands, methods based on the above approximation may smear out certain essential characteristics of bubbly liquids (see the Result sections for details).

**II B. Finite-amplitude method**

The distortion of acoustic waves is usually measured by the strength of the second harmonic, which can be obtained by applying the Fubini expansion to the nonlinear wave equation (Fubini 1935). Beyer introduced the finite-amplitude method briefly based on the Fubini expansion to the estimation of lossless media. Cobb improved the result by incorporating attenuation and dissipation, which had been presented before by Thuras for acoustic waves in air (Thuras, Jenkins et al. 1935). His method has been widely used in measuring the nonlinearity parameter of biological tissues and biomaterials (Zhu, Roos et al. 1983, Sehgal, Brown et al. 1986, Trarieux, Callé et al. 2014). In this method, the second harmonic pressure $p_2$ was given by the following relation (Cobb 1983)

$$|p_2| = \left(1 + \frac{B}{2A}\right) \frac{\left(e^{-2\alpha_1 x} - e^{-\alpha_2 x}\right)}{2\rho_0 c_0^3 (\alpha_2 - 2\alpha_1)} p_1^2$$

Here denote by $\Gamma = 1 + B/(2A)$ the nonlinearity coefficient (some authors also denote it by $\beta$). The first term "1" in this expression accounts for the self-convective effect contributing to the nonlinearity of propagation of acoustic waves arising from the continuity equation (Hamilton and Blackstock 1988, Everbach and Apfel 1995). Together with the inherent material nonlinearity parameter B/A, it describes the distortion of propagation waves in the medium. Experimental measurement can be done by measuring the values of the second harmonic ($p_2$) and fundamental ($p_1$) pressures, as well as the corresponding attenuation coefficients $\alpha_2$ and $\alpha_1$ at different locations away from the transducer source (by varying $x$). Then the nonlinearity parameter is determined by fitting the experimental data to the above equation (Cobb 1983, Wu and Zhu 1991, Varray, Basset et al. 2011). Later on, Ma developed a formula to evaluate the B/A of bubbly liquids by calculating the second harmonic with the help of a linear wave equation (Ma, Yu et al. 2004). In this formula, the nonlinearity is assumed to be mainly caused by nonlinear oscillations of the bubbles. Nonlinearity parameter of a suspension containing Albuneux® bubbles was estimated by Ma's formula and compared with experimental data from Wu (Wu, Zhu et al. 1995). However, Xia has shown that lipid-coated microbubbles oscillate almost linearly at the excitation pressure of 110 kPa (Xia, Porter et al. 2015). The Albuneux® bubbles, which are much stiffer than the typical lipid-coated microbubbles, are not expected to oscillate nonlinearly and generate a noticeable second harmonic at the excitation pressure of 22 kPa, a value used in their experiments. The same phenomenon has also been observed in the measurements of polymer-coated microbubbles (Xia, Paul et al. 2014). Therefore, the accuracy of the estimated B/A using the second harmonics is dubious. Furthermore, the dissipation of the medium may damp out the higher harmonics generated by the nonlinearity (Enflo and Hedberg 2006). In bubbly liquids, the dispersion of the mixture could also counteract the generation of the second harmonics resulting from the nonlinearity of the medium. The detailed discussion is presented in the Results section.

**II C. Mixture- law Method**

Since the presence of air bubbles in water can drastically alter the acoustic properties of the medium, Apfel developed a formula to calculate the B/A of gas-liquid mixtures (Apfel 1983, Everbach, Zhu et al. 1991). According to the theory, the compressibility of the gas phase is assumed to dominate the contributions to the altered acoustic characteristics. The effective B/A is given by

$$\frac{B}{A} = \frac{\left(\frac{\beta_0}{\rho_g c_g^2}\right)^2 \left(2+\frac{B}{A}\right)_g + \left(\frac{1-\beta_0}{\rho_0 c_0^2}\right)^2 \left(2+\frac{B}{A}\right)_w}{\left(\frac{\beta_0}{\rho_g c_g^2}+\frac{1-\beta_0}{\rho_0 c_0^2}\right)^2} - 2$$

Here the subscript *g* stands for gas phase and *w* stands for the water phase. $\beta_0$ is the volume fraction. Comparing with the finite-amplitude method that considers the second harmonics emissions due to nonlinear oscillations of bubbles, the above formula does not account for the oscillations of the bubbles explicitly. Therefore, the frequency-dependent nature of bubbly liquids and the effects of bubble coatings are not readily available in the method.

Based on the above analysis, to meet both practical and theoretical considerations, we may start with the linear dynamics of bubbles and directly employ the definition of B/A. Thus, our approach is still in the framework of the so-call thermodynamic method.

## III. NEW FORMULATION

### III A. Linear waves propagating in bubbly liquids

When a bubbly liquid contains sufficient dilute spherical microbubbles with a homogeneous distribution, we neglect the bubble-bubble interactions and multiple scatterings. The dispersion relation for linear waves propagating in the dilute bubbly liquid can be written in the form of (Commander and Prosperetti 1989)

$$\frac{1}{c^2} = \frac{1}{c_0^2} + 4\pi \int_0^{+\infty} \frac{1}{\omega_0^2} \frac{R_0}{1-\Omega^2 + j\delta\Omega} n(R_0) dR_0 \qquad (1)$$

where $c$ is the complex sound velocity in bubbly liquids, $\omega_0$ is the bubble's pulsation resonant frequency. Here the damping constant $\delta$ equals $2b/\omega_0$, and $b$ is the damping coefficient independent of $\omega_0$ according to the paper. Also, we have denoted $\Omega = \omega/\omega_0$. $n$ is the size distribution function of the initial radius $R_0$. The phase velocity $V_p$ thus can be calculated by $1/\text{Re}\{1/c\}$ while the attenuation coefficient $\alpha$ is given by $\omega \text{Im}\{1/c\}$. Eq.(1) suggests that the sound speed (phase velocity) in the bubbly liquid depends on the frequency of the propagating waves, as opposed to constant sound speed in the pure water.

The nonlinearity parameter is defined as (Beyer 1960)

$$\frac{B}{A} = 2\rho_0 c_0 (\frac{\partial V_p}{\partial P})_{s, P=P_0} \qquad (2)$$

where $P = P_0 + \Delta p$ is the instantaneous pressure, and $P_0$ is the hydrostatic pressure. Since in a small-pressure limit $\Delta P = \Delta P_0 + \Delta(\Delta p) \approx \Delta P_0$, taking the derivative of Eq.(1) with respect to $P_0$ and using the following relation

$$\frac{\partial \delta}{\partial P_0} = \frac{\delta}{\Omega} \frac{\partial \Omega}{\partial P_0} \tag{3}$$

we obtain

$$\frac{\partial c}{\partial P_0} = \frac{c^3}{c_0^3} \frac{\partial c_0}{\partial P_0} - \frac{4\pi c^3}{\omega^2} \int_0^{+\infty} \frac{\Omega}{\left(1 - \Omega^2 + j\delta\Omega\right)^2} \frac{\partial \Omega}{\partial P_0} R_0 n(R_0) dR_0 \tag{4}$$

It is not difficult to see that the derivative of $\Omega$ has the following form

$$\frac{\partial \Omega}{\partial P_0} = \omega \frac{\partial}{\partial P_0} \left(\frac{1}{\omega_0}\right) = -\frac{\omega}{\omega_0^2} \frac{\partial \omega_0}{\partial P_0} \tag{5}$$

and the derivative of the phase velocity $V_p$ can be written as

$$\frac{\partial V_p}{\partial P_0} = \text{Re}\{\frac{V_p^2}{c^2} \frac{\partial c}{\partial P_0}\} \tag{6}$$

By substituting Eq.(5) into Eq.(4) and combining the result with Eq.(2) and Eq.(6), we get

$$\frac{B}{A} = \text{Re}\{\frac{V_p^2}{c^2} \frac{c^3}{c_0^3} (2\rho_0 c_0) \frac{\partial c_0}{\partial P_0} + \frac{V_p^2}{c^2} (2\rho_0 c_0) \frac{4\pi c^3}{\omega^3} \int_0^{+\infty} \frac{\Omega^3}{\left(1 - \Omega^2 + j\delta\Omega\right)^2} \frac{\partial \omega_0}{\partial P_0} R_0 n(R_0) dR_0\} \tag{7}$$

To calculate the nonlinearity parameter, the derivative of the resonant frequency with respect to the hydrostatic pressure in the above equation has to be determined. This can be done by employing the linear dynamics of the bubble.

### III B. Linear oscillations of a microbubble

To derive the expression for $\partial \omega_0 / \partial P_0$, we consider the dynamics of a single ultrasound contrast agent bubble. Although the shell buckling phenomenon has been observed for UCAs with various shells, the occurrence of such phenomena in the linear region is still under debate. Predictions of coated microbubbles of linear oscillations by various models gave very similar results. More detailed analysis of this problem can be found in (Xia, Porter et al. 2015). Therefore, we employ the Constant Elasticity Model (CEM) as a matter of simplicity (Sarkar, Shi et al. 2005)

$$\rho_0\left(R\ddot{R}+\frac{3}{2}\dot{R}^2\right) = P_{G_0}\left(\frac{R_0}{R}\right)^{3\kappa} - 4\mu\frac{\dot{R}}{R} - \frac{4\kappa^s \dot{R}}{R^2} - \frac{2\gamma}{R} - \frac{2E^s}{R}[(\frac{R}{R_0})^2 - 1] - P_0 + P_A e^{-j\omega t} \quad (8)$$

where $R$ is the instantaneous radius of the bubble and $\dot{R} = dR/dt$, $\ddot{R} = d^2R/dt^2$, $P_{G_0}$ is the pressure inside the bubble, $\kappa$ is the polytropic constant, $\mu$ is the viscosity of the host liquid, $\gamma$ is a reference value of the interfacial tension, $\kappa^s$ and $E^s$ are the dilatational viscosity and elasticity of the bubble shell, respectively, and $P_A$ is the amplitude of the excitation pressure that has an angular frequency $\omega$. The dynamical equation for the bubble of a radius $R_0$, undergoing forced linear spherical pulsations ($R(t) = R_0 + X(t)$ and $|X(t)| << R_0$) at the external excitation pressure $P_A e^{-j\omega t}$ can be written in the form of

$$\ddot{X} + \frac{\dot{X}}{\rho R_0^2}(4\mu + \frac{4\kappa^s}{R_0}) + \frac{X}{\rho R_0^2}[3\kappa P_0 + \frac{4E^s}{R_0} + \frac{2\gamma}{R_0}(3\kappa - 1)] = \frac{P_A}{\rho R_0} e^{-j\omega t}$$

or (9)

$$\ddot{X} + \omega_0 \delta \dot{X} + \omega_0^2 X = F e^{-j\omega t}$$

where

$$\delta = \frac{1}{\rho_0 \omega_0 R_0^2}(4\mu + \frac{4\kappa^s}{R_0})$$

$$\omega_0 = \sqrt{\frac{1}{\rho_0 R_0^2}[3\kappa P_0 + \frac{4E^s}{R_0} + \frac{2\gamma}{R_0}(3\kappa - 1)]} \quad (10)$$

$$F = \frac{P_A}{\rho_0 R_0}$$

By substituting Eq.(10) into Eq.(5), we obtain

$$\frac{\partial \omega_0}{\partial P_0} = \frac{\partial}{\partial P_0}\sqrt{\frac{1}{\rho_0 R_0^2}[3\kappa P_0 + \frac{4E^s}{R_0} + \frac{2\gamma}{R_0}(3\kappa - 1)]} = \frac{1}{\omega_0}\frac{3\kappa}{2\rho_0 R_0^2} \quad (11)$$

By substituting Eq.(11) into Eq.(7), we finally have

$$\frac{B}{A} = \text{Re}\{\frac{V_p^2}{c^2}\frac{c^3}{c_0^3}(\frac{B}{A})_w + \frac{V_p^2}{c^2}\frac{12\kappa\pi c_0 c^3}{\omega^3}\int_0^{+\infty}\frac{1}{\omega_0}\frac{1}{R_0^2}\frac{\Omega^3}{(1-\Omega^2+j\delta\Omega)^2}R_0 n(R_0)dR_0\} \quad (12)$$

where

$$\left(\frac{B}{A}\right)_w = 2\rho_0 c_0 \frac{\partial c_0}{\partial P_0} \quad (13)$$

is the nonlinearity parameter of the pure water. It usually takes a value of 5 (Beyer 1960).

Eq.(12) is the first major result in this article. It correlates the nonlinearity parameter B/A to the dispersion relation, in which both the attenuation coefficient and phase velocity can be calculated. This equation captures the frequency-dependent nature of the B/A, which is neglected in the method of mixture law. In the meantime, it does not depend on the generation of the second harmonic that is required in the finite-amplitude method (Wu, Zhu et al. 1995).

**III C. Simplification for monodisperse bubbles**

Another significance of Eq.(12) is that one can easily evaluate the nonlinearity parameter B/A of a bubbly liquid as soon as the attenuation and phase velocity data are obtained in an acoustic measurement. To see this, we assume a monodisperse size distribution of the bubbles in the medium so that Eq.(1) becomes

$$\frac{1}{c^2} = \frac{1}{c_0^2} + 4\pi\frac{N}{\omega_0^2}\frac{R_0}{1-\Omega^2+j\delta\Omega} \quad (14)$$

where $N$ is the number of bubbles per unit volume. By substituting Eq.(14) into Eq.(12), we can obtain

$$\frac{B}{A} = \text{Re}\{\frac{V_p^2}{c^2}\left(\frac{c^3}{c_0^3}\left(\frac{B}{A}\right)_w + \frac{\kappa}{\beta_0}\frac{c^3}{c_0^3}\left(\frac{c_0^2}{c^2}-1\right)^2\right)\} \quad (15)$$

where $\beta_0 = 4\pi R_0^3 N/3$ is the bubble volume fraction. Evaluating of the left-hand side of Eq.(15) finally gives

$$\frac{B}{A} = \frac{V_p^2}{c_0^3} \mathrm{Re}\{c\} \left( 5 + \frac{\kappa}{\beta_0} \left( 1 - 2\frac{c_0^2}{|c|^2} + 4\frac{c_0^4}{|c|^4} \left( \frac{\mathrm{Re}\{c\}^2}{|c|^2} - \frac{3}{4} \right) \right) \right) \qquad (16)$$

The two expressions Re{c} and |c| can be deduced from Eq.(1)

$$\mathrm{Re}\{c\} = \frac{a^2 V_p}{\alpha^2 V_p^2 + a^2}$$

$$|c|^2 = \frac{a^2 V_p^2}{\alpha^2 V_p^2 + a^2} \qquad (17)$$

where $a = 20\omega \log_{10} e$. The attenuation coefficient is in dB per unit length.

Eq.(16) is the second major result in this article. It correlates the nonlinearity parameter B/A to two measurable experimental data: the attenuation coefficient $\alpha$ and the phase velocity $V_p$. Therefore, Eq.(16), along with Eq.(17), provides a convenient means for evaluating B/A of suspensions containing monodisperse free or coated microbubbles.

## IV. RESULTS AND DISCUSSIONS

### IV A. Comparisons

A recent result using a DEAT method that estimated the B/A of a suspension of polydisperse SonoVue microbubbles by gradually changing the hydrostatic pressure of the suspension is employed here for the purpose of comparison (Renaud, Bosch et al. 2015). Since these microbubbles are not monodisperse, we use Eq.(12) to calculate the nonlinearity parameter by employing the same shell parameters and size distribution of the SonoVue microbubbles that are described in the paper. The comparison is displayed in Figure 1, in which the curves have a fairly good agreement. It is known that the nonlinearity parameter is positive and frequency-independent in most biological materials (B/A = 5 ~ 10 ) (Everbach 1998). Figure 1 explicitly shows the frequency-dependent characteristic of the nonlinearity parameter in the bubbly liquid. It can also be negative. The maximum value of B/A is observed at the location close to the resonant frequency of the microbubbles (around 1.2 MHz) and it exceeds the values of common biological medium (5 - 10) by nearly two orders of magnitude.

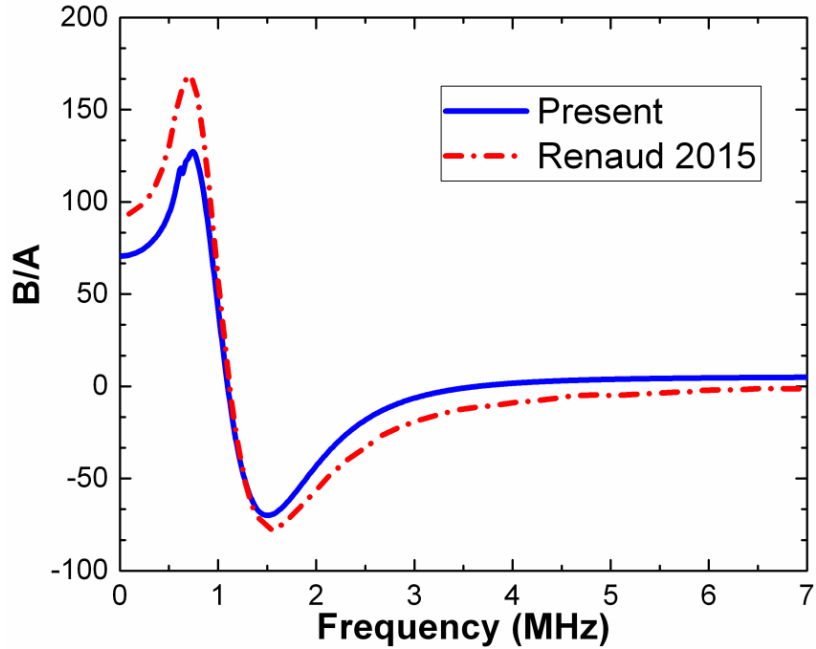

**Figure.1**: Comparison of the present calculation and the results from Renaud *et al.* 2015 using quasi-static approximations. The volume fraction is about $1\times10^{-6}$.

Experimentally measured nonlinearity parameters for suspensions containing monodisperse ultrasound contrast agents are rarely documented in the literature, here we only found experimental data for free monodisperse bubbles (bubbles without coatings) measured by the finite-amplitude method (Wu, Zhu et al. 1995, Ma, Yu et al. 2004). The B/A taken from their data, measured at the resonant frequency, is a single value for each bubble sample. For comparison, we thus took the maximum values in the frequency-dependent B/A curves plotted by Eq.(16). We put our calculations along with the published results in Table 1. We have taken $\kappa^s = 0, E^s = 0, \gamma = 0.072 \text{ N/m}$ in the CEM to represent the free bubbles.

**Table 1**: Comparison of the nonlinearity parameter B/A of bubbly liquids containing free monodisperse bubbles.

| Results | Bubbly liquids | B/A |
|---|---|---|
| Wu's Experimental data | $\beta_0 = 8\times10^{-3}, R_0 = 3.3 \,\mu\text{m}$ | 2x10$^5$ |

| Wu's Calculation | $\beta_0 = 8 \times 10^{-3}$, $R_0 = 3.3$ μm | 5 x$10^5$ |
| --- | --- | --- |
| Ma's Calculation | $\beta_0 = 8 \times 10^{-3}$, $R_0 = 3.3$ μm | 4.88x$10^5$ |
| Present Calculation | $\beta_0 = 8 \times 10^{-3}$, $R_0 = 3.3$ μm | 4.86x$10^5$ |
| Wu's Experimental data | $\beta_0 = 9 \times 10^{-4}$, $R_0 = 2.4$ μm | 9x$10^4$ |
| Wu's Calculation | $\beta_0 = 9 \times 10^{-4}$, $R_0 = 2.4$ μm | 3 x$10^5$ |
| Present Calculation | $\beta_0 = 9 \times 10^{-4}$, $R_0 = 2.4$ μm | 4.75x$10^5$ |

For the bubble sample of radius 3.3 μm, all the calculated values of B/A are similar and almost 2.5 times higher than the experimental data; whereas for the bubble sample of radius 2.4 μm, the present calculation is larger than that of Wu's calculation and experiment. Wu suggested the value of B/A for bubbly liquids containing free bubbles are about $10^4 \sim 10^5$. All the results from the theoretical calculations are fallen in this range; however, they are higher than the experimental results. The discrepancy between the experimental and theoretical results may be due to the high volume fractions used in their experiments. This is because the measured B/A at the resonant frequency of the bubble does not correspond to the maxima in the B/A curve when the bubble volume fraction goes large values. To see this, we plot the phase velocity, attenuation, and B/A curves vs frequency for the bubble sample of radius 3.3 μm in Figure 2. For the monodisperse bubbles, the resonant frequency can be identified by the location of the maximum attenuation coefficient in the second panel of each sub-figure, which is 1.15 MHz. The third panel in Figure 2a shows that the maximum B/A locates at 11.13 MHz for the bubbly liquid with $\beta_0 = 8 \times 10^{-3}$ in our calculation, far away from the resonant frequency of the bubble (1.15 MHz). When the volume fraction is decreasing, the third panel in Figure 2b, 2c, or 2d shows that the location of the maximum B/A approaches the resonant frequency. Meanwhile, the variation of the phase velocity and attenuation in Figure 2 suggests that the bubbly liquid is a highly dispersive medium with strong dissipations at large volume fractions, which may also explain the occurrence of the discrepancy as the dispersion and attenuation in the medium counteract the nonlinear distortion and thus decreases the measured values of B/A.

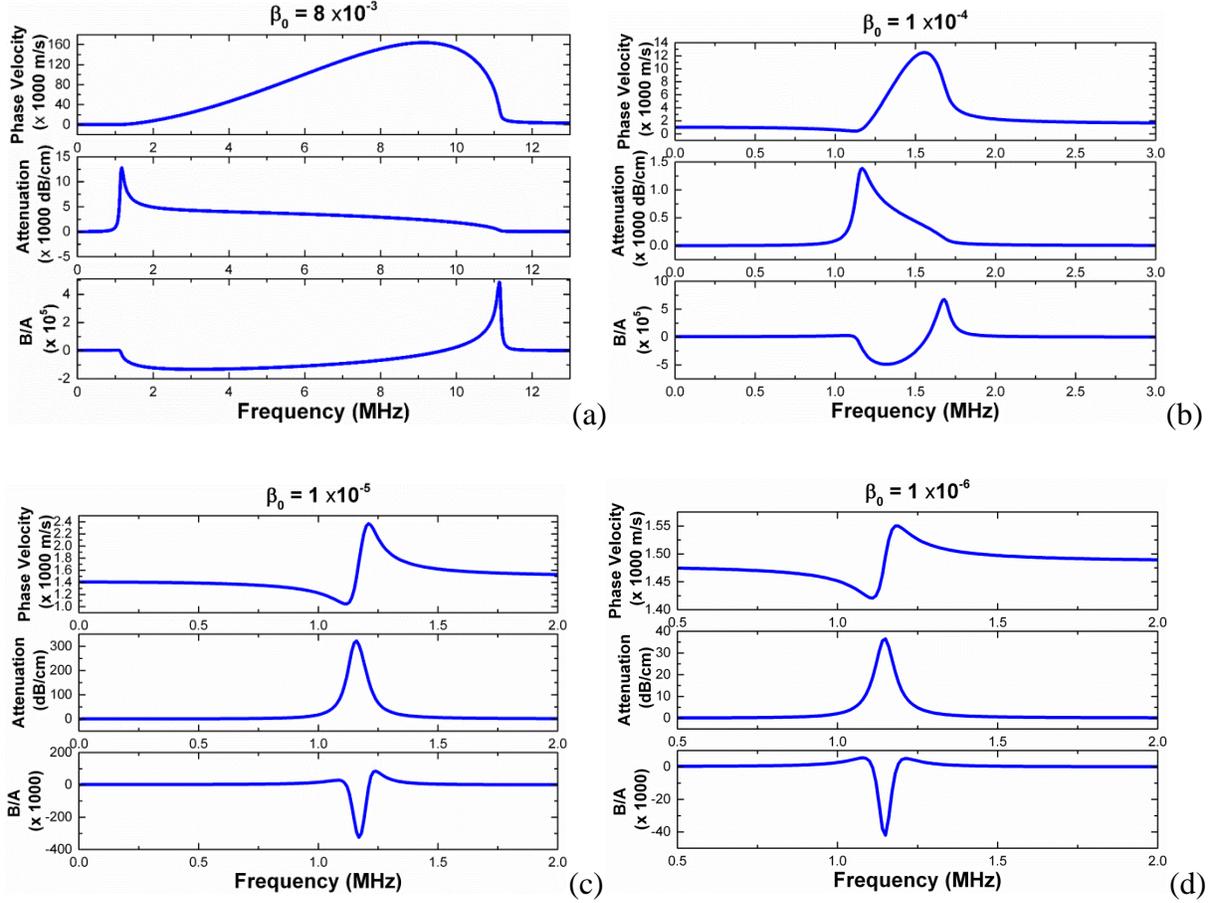

**Figure.2**: Calculations of the nonlinearity parameter B/A, attenuation, and phase velocity at the bubble volume fraction of $8\times10^{-3}$ (a), $1\times10^{-4}$ (b), $1\times10^{-5}$ (c), and $1\times10^{-6}$ (d) for bubbly liquids containing free monodisperse bubbles.

### IV B. Negative nonlinearity parameter

In studying shock waves, the nonlinear coefficient $\Gamma$ (also called fundamental derivative) plays a key role in the distortion of pressure and shock waves (Kluwick 1991). A negative $\Gamma$ usually results in many unusual phenomena that cannot be explained by the theory of classical nonlinear acoustics (Thompson, Carofano et al. 1986, Thompson 1991, Guardone, Colonna et al. 2014). Experimental observations on the negative nonlinearity parameter of bubbly liquids or suspension of air-filled glass beads were reported by many researchers (Trivett, Pincon et al. 2006, Hélène, Anthony et al. 2009, Zaitsev, Dyskin et al. 2009, Trarieux, Callé et al. 2014, Renaud, Bosch et al. 2015). Zaitsev et al and Renaud et al. claimed that the negative values of B/A of a suspension containing coated bubbles were induced by the shell buckling of the bubbles in the suspensions, which results in the decrease of the phase velocity with increasing the hydrostatic pressure. However, as we mentioned in the above literature, negative nonlinearity parameters were also reported for liquid-vapor systems, in which the buckling phenomenon may not exist. Furthermore, the present model that does not assume the state of shell buckling still predicts the existence of negative values. In other words, although the buckling of the shell could have effects on the negative B/A, it is not the trigger for the occurrence of negative values.

Eq.(16) suggests that the occurrence of negative values of B/A is determined by the complex sound speed, and B/A can change signs with respect to the frequency of propagating waves. This suggests that the presence of bubbles can significantly change the thermodynamic properties of the host medium, and bubbly liquids are possibly retrograde fluids (Cramer and Sen 1986). Since the complex sound speed is determined by the dispersion relation Eq.(14), we investigate the effects of bubble volume fraction $\beta_0$, initial radius $R_0$, hydrostatic pressure $P_0$, and excitation frequency $f$ (the frequency of propagating waves) separately. The curves plotted in Figure 3 are taken from the minimum B/A in each of the range. In Figure 3a, we keep $R_0 = 3.3\,\mu\text{m}$, $f = f_0$, and $P_0 = 101\,\text{kPa}$. B/A decreases smoothly with increasing $\beta_0$. Since the maximum and minimum B/A always locate around the resonant frequencies of bubbles, it is not difficult to figure out that the dynamics of the bubbles contribute to the great magnitude of the B/A. In Figure 3b, we keep other parameters constant ($\beta_0 = 10^{-6}$, $f = f_0$, and $P_0 = 101\,\text{kPa}$) and vary the bubble radius, the negative B/A decreases smoothly until $R_0 = 3.78\,\mu\text{m}$ where the corresponding resonant frequency is 0.9839 MHz, then it increases drastically and reaches a constant B/A = 5 at $R_0 = 3.84\,\mu\text{m}$ where the corresponding resonant frequency is 0.9669 MHz. Figure 3b indicates that even a small change in bubble radius could have a significant impact on the negative B/A. The effect of hydrostatic pressure on the B/A was investigated experimentally for a composite medium consisting of hollow microspheres suspended in Castor oil (Trivett, Pincon et al. 2006). In Figure 3c, we generate a curve of similar shape (here $R_0 = 3.3\,\mu\text{m}$, $f = f_0$, and $\beta_0 = 10^{-6}$) to the experimental result. Varying the hydrostatic pressure in the range of 0-60 kPa does not change the values of B/A, which remains 5 (the resonant frequency is 0.9544 MHz at 60 kPa). B/A decreases fast in the range of 60-70 kPa, touches the bottom at 70 kPa, at which the resonant frequency is 1.0043 MHz, and then increases smoothly. The above observations suggest negative B/A depends on the input parameters nonlinearly. A subtle change in the above parameters could result in significant variations in B/A.

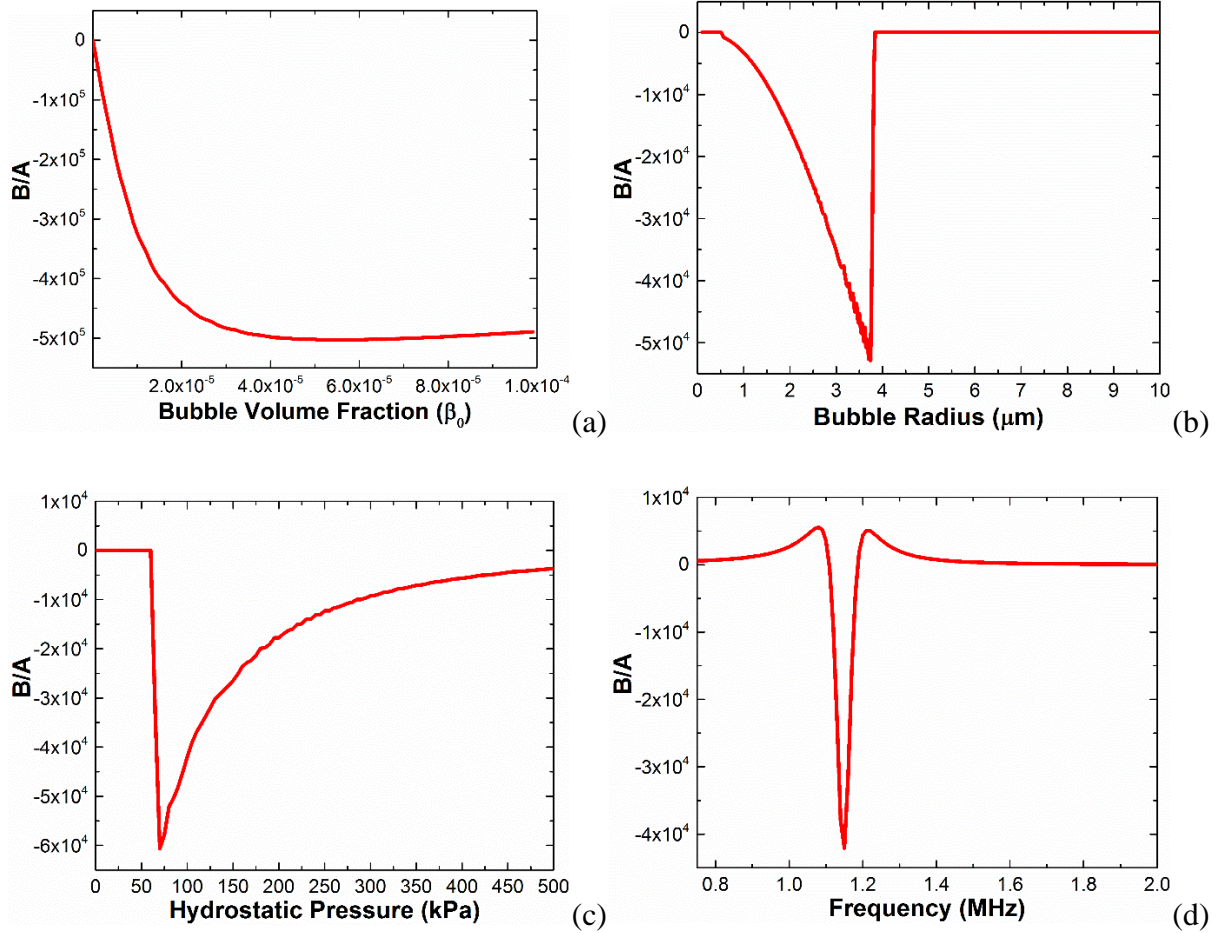

**Figure.3**: Calculations of the negative nonlinearity parameter B/A *vs.* bubble volume fraction (a), bubble radius (b), hydrostatic pressure (c), and excitation frequency (d).

The dependence of nonlinearity parameter B/A on the excitation frequency was found in several experimental studies (Hélène, Anthony et al. 2009, Renaud, Bosch et al. 2015). A comparison has been given in Figure 1. Figure 3d shows that the B/A is 5 at zero frequency, and the extremum magnitudes of B/A occur near the resonant frequency. When the frequency further increases, the B/A decreases to 5. Therefore, the resonance of the bubbles contributes to the large values of B/A of the bubbly liquids.

**IV C. Effects of bubble volume fractions**

To further investigate the nonlinearity parameter of bubbly liquids containing free microbubbles with a monodisperse distribution (avoiding debates on the buckling effect), we plot the variation of B/A with respect to the bubble volume fraction in Figure 5a, in which the results of the present paper and the mixture law are represented by the solid and dashed curves, respectively. In the case of the present calculation (the solid curve), the value of B/A equals 5 when the bubble volume fraction $\beta_0$ is zero, which is the nonlinearity parameter of the pure water. B/A does not vary too much and remains below 10 until the volume fraction reaches at $10^{-9}$, indicating the concentration of bubbles has little impact on the overall nonlinearities when the bubble volume fraction is low (Bjørnø 1982). With continuing increases in the volume fraction, the present calculation shows a peak of B/A $= 7.92 \times 10^5$ at the location of $2.48 \times 10^{-4}$. After the peaks, the B/A decreases with further increasing the volume fraction, which lowers down to 328.44 at $\beta_0 = 1$. In the case of the mixture law, the estimated curve (the dashed curve) has a similar trend to that of the blue curve. However, the maximum B/A estimated by the mixture law is about $8.98 \times 10^3$ (at the location of $8.81 \times 10^{-5}$), which is much lower than the present calculation. This discrepancy is the result of missing the dynamic effect of the bubbles in the method of mixture law. Although the mixture law of nonlinearity parameter can accurately calculate the effective B/A of some mixtures, it is not able to capture the contributions from the oscillations of the bubbles in the bubbly liquid, particularly for the bubbles at the resonance.

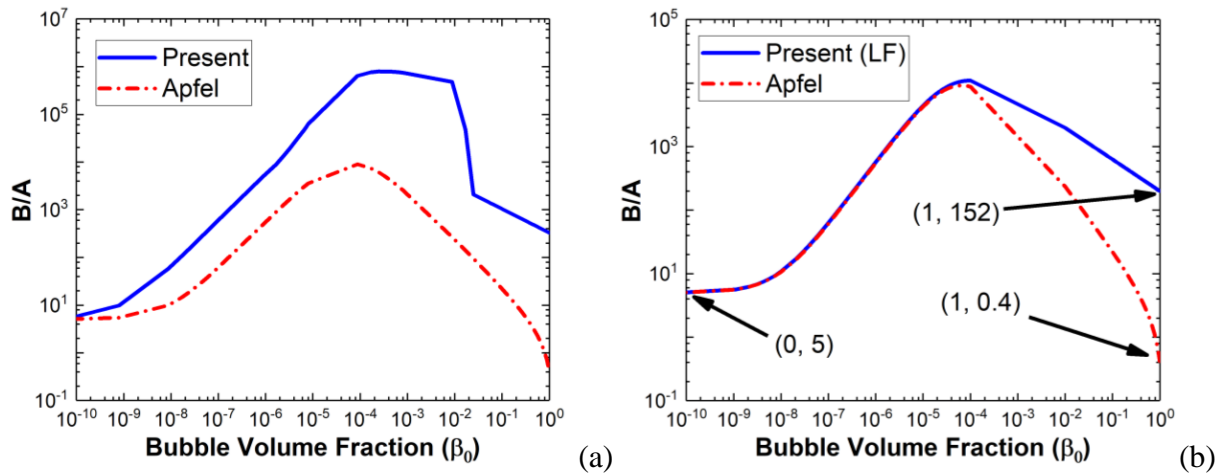

**Figure.4**: The calculated nonlinearity parameter B/A of bubbly liquids containing free monodisperse bubbles *vs.* Bubble volume fraction. (a): present calculation (blue curve, the excitation frequency is equal to the resonant frequency) and Calculation using Apfel's formula

(red dashed curve); (b): present calculation in low-frequency limit (blue curve, the excitation frequency is below the resonant frequency) and Calculation using Apfel's formula (red dashed curve).

To directly compare the results of the two methods, we evaluate the B/A at a low-frequency limit, where the impact of bubble dynamics can be neglected. In Figure 4b, the two calculations overlap in the region $\beta_0 < 10^{-4}$, with the peaks locating at $2.48 \times 10^{-4}$. With further increasing the volume fraction, the present calculation decreases to 152 at $\beta_0 = 1$. Note that for ideal gasses, $B/A = \kappa - 1$ (Fox and Wallace 1954, Beyer 1960) and thus the upper bound of B/A, according to the formula, should be 1.4 - 1= 0.4 (at which $\beta_0 = 1$, see the red dashed curve in Figure 4b), instead of 152. This discrepancy is due to the assumption of the effective medium approach for deriving the dispersion relation Eq.(1). In the derivation, we have approximated the density of bubbly liquids $\rho_m = (1-\beta_0)\rho_0 + \beta_0 \rho_g \approx (1-\beta_0)\rho_0$, and our formula is not able to capture bubbly liquids with larger volume fractions, especially when $\beta_0$ approaches 1. Thus, for the accurate estimation of B/A with the present method, the value of bubble volume fraction is suggested to be less than $10^{-4}$. When $\beta_0 > 10^{-4}$, we may need to employ the theory that includes higher-order multiple scattering effects (Kargl 2002).

**IV D. Effects of bubble encapsulations**

To investigate the effects of the outer shells on coated bubbles in a bubbly liquid, we plotted the B/A with respect to the dilatational viscosity and dilatational elasticity in Figure 5. The bubble volume fraction ($\beta_0 = 9.0 \times 10^{-4}$) and surface tension ($\gamma = 0.02 \text{ N/m}$) were kept constant in the calculations. Figure 5 shows that the nonlinearity parameter B/A decreases with increasing shell parameter (viscosity or elasticity), as the shells damp the oscillations of the bubbles.

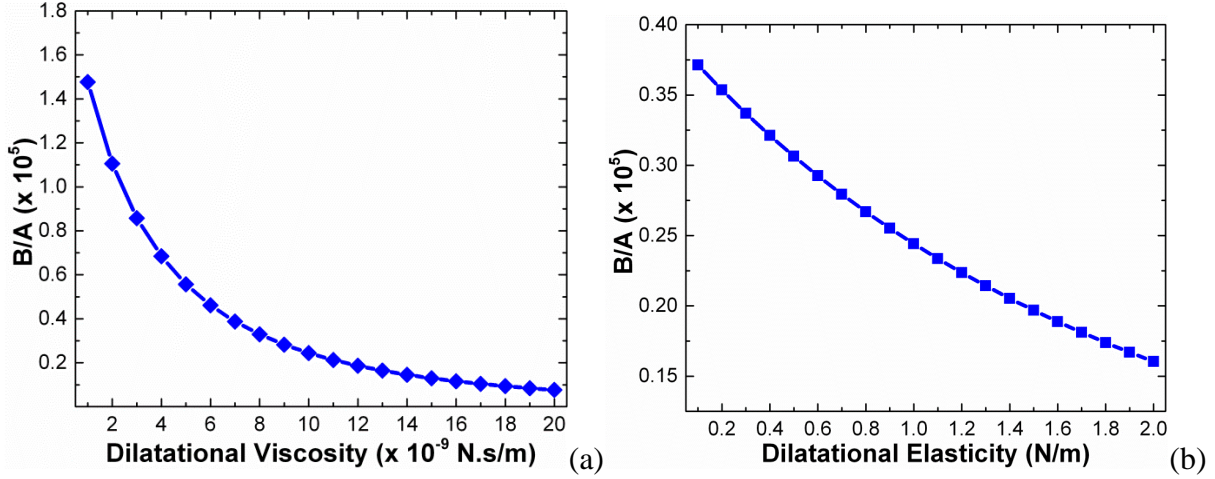

**Figure.5**: The calculated nonlinearity parameter B/A *vs.* bubble's dilatational viscosity (a). ( $\beta_0 = 9.0 \times 10^{-4}$, $R_0 = 2.4\,\mu m$, and $E^s = 1.0\,N/m$ ); bubble's dilatational elasticity (b). ($\beta_0 = 9.0 \times 10^{-4}$, $R_0 = 2.4\,\mu m$, and $\kappa^s = 1.0 \times 10^{-8}\,N.s/m$).

## IV E. Concerns on the assumption of linear propagating waves

For bubbly liquids, whether a high nonlinearity parameter gives rise to significant distortion of propagating waves can be examed by the Goldberg number $\epsilon = 1/(\alpha l)$ (Goldberg 1962, Blackstock 1964), here $l = c_0^2/(\Gamma U_0 \omega_0)$ is the discontinuity distance (at which a shock forms), and $U_0$ is the initial particle velocity (Trivett, Pincon et al. 2006). If the Goldberg number is smaller than 1, the linear attenuation dominates, and distortion of the wave profile is not evident. Here we assume $U_0 = P/(\rho_0 c_0)$. In the case of the SonoVue microbubbles, we can take $P = 50\,kPa$, $\omega_0 = 2\,MHz$, $\Gamma = 200$, and $\alpha = 5\,dB/cm$, then the Goldberg number is $\epsilon = 0.012$, which is much smaller than 1. This indicates that the distortion is not significant, and the propagating wave in the suspension of the SonoVue microbubbles is still linear, such that the linear attenuation theory is valid. Similar results for the bubbly liquids discussed in Figure 2 can also be obtained. Meanwhile, the dispersion of the bubbly liquids also indicates that nonlinear effects are of no general significance (Crighton 1991). Therefore, a highly nonlinear medium (e.g., bubbly liquids) does not necessarily distort propagating waves.

## V. CONCLUSIONS

We have developed a method for evaluating the nonlinearity parameter B/A of suspensions of ultrasound contrast agents by using the theory of linear acoustic waves in the media. The frequency-dependent B/A of a suspension containing polydisperse SonoVue microbubbles was estimated by Eq.(12), which is in agreement with results obtained from the dynamic acousto-elastic testing. Negative values of B/A were found related to the complex sound speed in bubbly liquids. Large magnitudes of B/A of suspensions containing free monodisperse bubbles that had been documented in both experimental reports and the finite-amplitude method were reproduced by Eq.(16). It shows that maximum values of B/A always locate near the resonant frequency of the bubbles when the bubble volume fraction $\beta_0 < 10^{-4}$. The method also demonstrated that the nonlinearity parameter B/A of a suspension of ultrasound contrast agents decreases with increasing the shell viscosity and elasticity of the coated microbubbles. In the low-frequency limit, the variation of B/A with respect to the volume fraction of bubbles was calculated by the present method and showed great similarity to that estimated by the method of mixture law. The present method suggests that the resonant oscillation of bubbles, together with the bubble volume fraction, contributes to the large magnitude of the nonlinearity parameter. Meanwhile, Eq.(16), along with Eq.(17), provides an explicit formula for estimating B/A from two experimentally measurable variables: the attenuation and phase velocity. Note that we adopted the effective medium assumption (low bubble volume fraction) with the restriction to the linear dynamics of bubbles, which is sufficient for acoustic experiments of linear attenuation, future works can be focused on bubbly liquids of large bubble volume fractions, *e.g.* $\beta_0 > 10^{-4}$.

## ACKNOWLEDGMENTS

L. Xia thanks Dr. Guillaume Renaud at French National Centre for Scientific Research, Paris, for he generously sharing the size distribution data of the SonoVue bubbles. L. Xia thanks Prof. Kausik Sarkar at The George Washington University for helpful discussions. L. Xia appreciates the anonymous reviewer for her/his careful readings and valuable comments.